\documentstyle[12pt,epsf]{article}

\newcommand{\MS}{$\overline{\mathrm{MS}}$\ }
\newcommand{\LMS}{\Lambda_{\overline{\mathrm{MS}}}}
\newcommand{\ep}{\varepsilon}

\begin{document}

\begin{titlepage}

\begin{flushright}
CERN-TH/97-102\\
Budker INP 97-60\\
hep-ph/9707318
\end{flushright}

\vspace{1cm}
\begin{center}
\Large\bf
Higher-order estimates of the chromomagnetic\\
moment of a heavy quark
\end{center}

\vspace{0.5cm}
\begin{center}
A.G. Grozin\\[0.1cm]
{\sl Budker Institute of Nuclear Physics, Novosibirsk 630090, 
Russia}\\[0.3cm]
M. Neubert\\[0.1cm]
{\sl Theory Division, CERN, CH-1211 Geneva 23, Switzerland}
\end{center}

\vspace{1cm}
\begin{abstract}
\vspace{0.2cm}\noindent
The leading $\beta_0^{n-1}\alpha_s^n$ terms in the Wilson coefficient
and anomalous dimension of the chromomagnetic operator in the
heavy-quark effective Lagrangian are summed to all orders of
perturbation theory. The perturbation series for the anomalous
dimension is well behaved, while that for the Wilson coefficient
exhibits a divergent behaviour already in low orders, caused by a
nearby infrared renormalon singularity. The resulting ambiguity is
commensurate with terms of order $1/m^2$ in the effective Lagrangian,
whose corresponding ultraviolet renormalons are identified. An
excellent approximation for the scheme-invariant Wilson coefficient at
next-to-next-to-leading order in renormalization-group improved
perturbation theory is obtained.
\end{abstract}

\vspace{1cm}
\centerline{(Submitted to Nuclear Physics B)}

\vfil
\noindent
CERN-TH/97-102\\
July 1997

\end{titlepage}

\section{Introduction}

The properties of hadronic bound states containing heavy quarks are
characterized by a large separation of energy scales. Effects
associated with the heavy-quark mass $m$ are perturbative and can be
controlled once they have been separated from other, long-distance
effects. This separation is most conveniently done using an effective
low-energy theory. In the case of hadrons containing a single heavy
quark, the relevant effective theory is the heavy-quark effective
theory (HQET) (see~\cite{review} for a review). Its Lagrangian is
\begin{equation}
   {\cal L} = \bar h_v\,i v\!\cdot\!D\,h_v
   + \frac{1}{2m}\,\bar h_v(i D_\perp)^2 h_v
   + \frac{C(m,\mu)}{4m}\,
   \bar h_v\,g_s\sigma_{\mu\nu} G^{\mu\nu} h_v + O(1/m^2) \,,
\label{Leff}
\end{equation}
where $v$ is the velocity of the hadron containing the heavy quark,
$h_v$ the velocity-dependent heavy-quark field, and $D_\perp$ the
covariant derivative orthogonal to $v$. The operators arising at order
$1/m$ in the effective Lagrangian correspond to the kinetic energy of
the heavy quark and its chromomagnetic interaction~\cite{EiHi,FGL}.
Hadronic matrix elements of these operators appear in many applications
of HQET. The coefficient of the kinetic-energy operator is fixed by
reparametrization invariance, an invariance under infinitesimal changes
of the velocity~\cite{LuMa}. Hence, the only non-trivial short-distance
coefficient in the effective Lagrangian at next-to-leading order in
$1/m$ is the coefficient $C(m,\mu)$ of the chromomagnetic operator.

The dependence of the Wilson coefficient on the large scale $m$ can be
factorized using the renormalization group. In general,
\begin{equation}
   C(m,\mu) = C(m,m)\,\exp\!\int\limits_{\alpha_s(\mu)}^{\alpha_s(m)}
   d\alpha_s\,\frac{\gamma(\alpha_s)}{\beta(\alpha_s)}
   \equiv \widehat C(m)\,K(\mu) \,,
\label{RGEsol}
\end{equation}
where
\begin{equation}
   \beta(\alpha_s) = - \frac{d\alpha_s}{d\ln\mu}
   = 2 \alpha_s \left[ \beta_0\,\frac{\alpha_s}{4\pi}
   + \beta_1 \left(\frac{\alpha_s}{4\pi}\right)^2
   + \dots \right] \,, \qquad
   \beta_0 = \frac{11}{3} C_A - \frac{4}{3}\,T_F n_f
\label{beta}
\end{equation}
is the $\beta$-function, and
\begin{equation}
   \gamma(\alpha_s) = \gamma_0 \frac{\alpha_s}{4\pi}
   + \gamma_1 \left(\frac{\alpha_s}{4\pi}\right)^2 + \dots
   = \frac{C_A\alpha_s}{2\pi} \left[ 1
   + \frac{13\beta_0 - 25 C_A}{6}\,\frac{\alpha_s}{4\pi}
   + \dots \right]
\label{gamma2}
\end{equation}
the anomalous dimension of the chromomagnetic operator (given in the
\MS scheme). The two-loop coefficient $\gamma_1$ of the anomalous
dimension was calculated in~\cite{ABN,CG}, and the two-loop initial
condition $C(m,m)$ in~\cite{CG}. The corresponding one-loop expressions
were obtained much earlier in~\cite{EiHi}. The factor $K(\mu)$
in~(\ref{RGEsol}) compensates the scheme and scale dependence of the
chromomagnetic operator in the effective Lagrangian~(\ref{Leff}). All
short-distance physics relevant to physical observables is contained in
the renormalization-scheme invariant function $\widehat C(m)$ which, up
to an $m$-independent normalization, is given by
\begin{equation}
   \widehat C(m) = \big[ \alpha_s(m) \big]^{\gamma_0/2\beta_0}\,
   \Big[ 1 + c(m) \Big] \,, \qquad
   c(m) = \sum_{n=1}^\infty c_n \left( \frac{\alpha_s(m)}{4\pi}
   \right)^n \,.
\label{hatC}
\end{equation}
In the \MS scheme, the next-to-leading correction is given by
\cite{ABN}
\begin{equation}
   c_1 = \left( \frac{25}{6} C_A + 2 C_F \right)
   - \left( \frac{55}{6} C_A + 3 C_F \right) \frac{C_A}{\beta_0}
   + \left( 7 C_A + 11 C_F \right) \frac{C_A^2}{\beta_0^2} \,.
\label{c1}
\end{equation}
In the above equations, $C_F=\frac 12(N-1/N)$, $C_A=N$, and $T_F=1/2$
are the colour factors for an SU$(N)$ gauge group.

As an application, consider the mass splitting between the ground-state
pseudoscalar and vector mesons containing a heavy quark. Its leading
contribution comes from the chromomagnetic operator in the effective
Lagrangian, which is the first term breaking the heavy-quark spin
symmetry. Including terms of order $1/m^2$, one
obtains~\cite{Bigi}\footnote{A similar mass formula has been derived in
\protect\cite{Mannel}; however, the term proportional to $\rho_A^3$ is
missing there.}
\begin{eqnarray}
   &&M_V - M_P = \frac{2C(m,\mu)}{3m}\,\mu_G^2(\mu)
    \nonumber\\
   &&\quad\mbox{}+ \frac{1}{3m^2} \bigg[
    C(m,\mu)\,\rho_{\pi G}^3(\mu) + C^2(m,\mu)\,\rho_A^3(\mu)
    - C_{LS}(m,\mu)\,\rho_{LS}^3(\mu) \bigg] \,,
\label{split}
\end{eqnarray}
where $\mu_G^2(\mu)$ is the matrix element of the chromomagnetic
operator between ground-state mesons. Similarly, the hadronic
parameters $\rho_i^3(\mu)$ are defined in terms of the matrix elements
of operators appearing at order $1/m^2$ in the effective Lagrangian
\cite{Bigi}: $\rho_{LS}^3(\mu)$ parametrizes the spin-orbit
interaction, while $\rho_{\pi G}^3(\mu)$ and $\rho_A^3(\mu)$ are
bilocal matrix elements of kinetic-chromomagnetic and
chromomagnetic-chromomagnetic insertions. (We omit the contributions
from 4-quark operators \cite{Korn}, which are irrelevant to our
discussion.) These parameters are independent of the heavy-quark mass,
but they depend on the renormalization scale in such a way that the
scale dependence of the Wilson coefficients is compensated. The
coefficient of the spin-orbit term is related by reparametrization
invariance to that of the chromomagnetic operator \cite{CKO}--\cite{FM}
(see also the appendix of \cite{Korn}): $C_{LS}(m,\mu)=2 C(m,\mu)-1$.
Introducing then the renormalization-scheme invariant parameters
\begin{eqnarray}
   \widehat\mu_G^2 &=& K(\mu)\,\mu_G^2(\mu) \,, \nonumber\\
   \widehat\rho_{\pi G}^3 &=& K(\mu)\,\rho_{\pi G}^3(\mu)
    + 2 [1-K(\mu)]\,\rho_{LS}^3(\mu) \,, \nonumber\\
   \widehat\rho_A^3 &=& K^2(\mu)\,\rho_A^3(\mu) \,, \qquad
   \widehat\rho_{LS}^3 = \rho_{LS}^3(\mu) \,,
\end{eqnarray}
we find
\begin{equation}
   M_V - M_P = \frac{2\widehat C(m)}{3m}\,\widehat\mu_G^2
   + \frac{1}{3m^2} \left[ \widehat C(m)\,(\widehat\rho_{\pi G}^3
   - 2\widehat\rho_{LS}^3) + \widehat C^2(m)\,\widehat\rho_A^3
   + \widehat\rho_{LS}^3 \right] \,.
\label{MVMP}
\end{equation}
All short-distance effects in this relation are contained in the single
coefficient $\widehat C(m)$, which is the object of our study.

\boldmath
\section{All-order results in the large-$\beta_0$ limit}
\unboldmath

The perturbation series for $C(m,\mu)$ can be arranged as
\begin{equation}
   C(m,\mu) = 1 + \sum_{L=1}^\infty \sum_{n=0}^{L-1}\,
   a_{n}^{(L)}(m/\mu)\,\beta_0^n\,\alpha_s^L(\mu) \,,
\label{Series}
\end{equation}
where $L$ is the number of loops, and $\beta_0$ the leading coefficient
of the $\beta$-function in~(\ref{beta}). In this letter, we sum the
terms of order $\beta_0^{L-1}\alpha_s^L$ to all orders of perturbation
theory. In other words, we consider the limit of large $\beta_0$ for
fixed $\beta_0\alpha_s$ and calculate the coefficient $C(m,\mu)$ to
order $1/\beta_0$, neglecting terms of order $1/\beta_0^2$ and higher.
Strictly speaking, there is no sensible limit of QCD in which $\beta_0$
may be considered a large parameter (except, maybe, $n_f\to-\infty$);
however, retaining only the leading $\beta_0$ terms often gives a good
approximation to exact multi-loop results (see, e.g., \cite{BG}), in
particular in cases when there is a nearby infrared renormalon
\cite{scale}. At the least, it will provide us with some information
about the summability of the perturbation series.

The coefficients $a_{L-1}^{(L)}$ of the terms with the highest degree
of $\beta_0$ in~(\ref{Series}) are determined by diagrams with $L-1$
light-quark loops, which are rather straightforward to calculate. We
work in dimensional regularization with $d=4-2\ep$ space-time
dimensions and adopt the \MS subtraction scheme. At first order in
$1/\beta_0$, multiplicative renormalization simply amounts to a
subtraction of the $1/\ep^n$ poles, and coupling-constant
renormalization is given by ($\bar\mu^2=\mu^2 e^\gamma/4\pi$)
\begin{equation}
   \frac{\beta_0 g_0^2}{(4\pi)^2} = \bar{\mu}^{2\ep}\,
   \frac{b}{1+b/\ep} \,, \qquad
   b = \frac{\beta_0\alpha_s(\mu)}{4\pi}
   = \frac{1}{2\ln(\mu/\LMS)} \,.
\label{Renorm}
\end{equation}
The perturbation series for $C(m,\mu)$ can then be written as
\begin{equation}
   C(m,\mu) = 1 + \frac{1}{\beta_0}\,\sum_{L=1}^{\infty}\,
   \frac{F(\ep,L\ep)}{L} \left( \frac{b}{\ep+b} \right)^L
   - \hbox{(minimal subtractions)} + O(1/\beta_0^2) \,.
\label{Struct}
\end{equation}
The function $F(\ep,u)$ is regular at $\ep=u=0$. Following the methods
of~\cite{meth1,meth2} (used also in~\cite{BG}), we now expand
$F(\ep,u)$ in powers of $\ep$ and $u$, and $[b/(\ep+b)]^L$ in powers of
$b/\ep$, to obtain a quadruple sum in (\ref{Struct}). Combinatoric
identities relate the $1/\ep$ terms, and hence the anomalous dimension
of the chromomagnetic operator, to the Taylor coefficients of
$F(\ep,0)$~\cite{meth1}:
\begin{equation}
   \gamma = \frac{2b}{\beta_0}\,F(-b,0) + O(1/\beta_0^2) \,.
\label{gamma}
\end{equation}
The finite terms, which determine the Wilson coefficient itself,
receive contributions from the Taylor coefficients of both $F(\ep,0)$
and $F(0,u)$~\cite{meth2}:
\begin{eqnarray}
   C(m,\mu) &=& 1 + \frac{1}{\beta_0} \int\limits_{-b}^0\!
    d\ep\,\frac{F(\ep,0)-F(0,0)}{\ep} \nonumber\\
   &&\mbox{}+ \frac{1}{\beta_0} \int\limits_0^\infty\!du\,
    e^{-u/b}\,\frac{F(0,u)-F(0,0)}{u} + O(1/\beta_0^2) \,.
\label{finite}
\end{eqnarray}
In this expression, all dependence on the heavy-quark mass resides in
the function $F(0,u)\sim (\mu/m)^{2u}$, while a dependence on the
renormalization scale also enters through the coupling
$b\sim\alpha_s(\mu)$. By separating in~(\ref{finite}) terms depending
on the two scales $m$ and $\mu$, we find that the next-to-leading
logarithmic correction $c(m)$ in~(\ref{hatC}) is given by
\begin{eqnarray}
   c(m) &=& \frac{1}{\beta_0} \int\limits_0^\infty\!du\,
    e^{-u[1/b(m)+\kappa]}\,S(u) + O(1/\beta_0^2) \,, \nonumber\\
   S(u) &=& e^{\kappa u} \left. \frac{F(0,u)-F(0,0)}{u}
    \right|_{\mu=m} \,.
\label{delc}
\end{eqnarray}
The same result can also be derived by evaluating~(\ref{RGEsol}) in the
large-$\beta_0$ limit. Here $b(m)=\beta_0\alpha_s(m)/4\pi$, and the
constant $\kappa$ is introduced to compensate the scheme dependence of
$1/b(m)$. The most natural choice is to have $\kappa=-5/3$ in the \MS
scheme. Then the combination
\begin{equation}
   \frac{1}{b(m)} + \kappa = 2 \ln(m/\LMS) - \frac 53
   \equiv 2 \ln(m/\Lambda_{\mathrm{V}})
\label{aV}
\end{equation}
defines the inverse coupling in the so-called V scheme~\cite{BLM}, with
$\Lambda_{\mathrm{V}}=e^{5/6}\LMS$. The function $S(u)$ is the Borel
transform of the perturbation series for $c(m)$ in that scheme.

\begin{figure}
\epsfxsize=12cm
\centerline{\epsffile{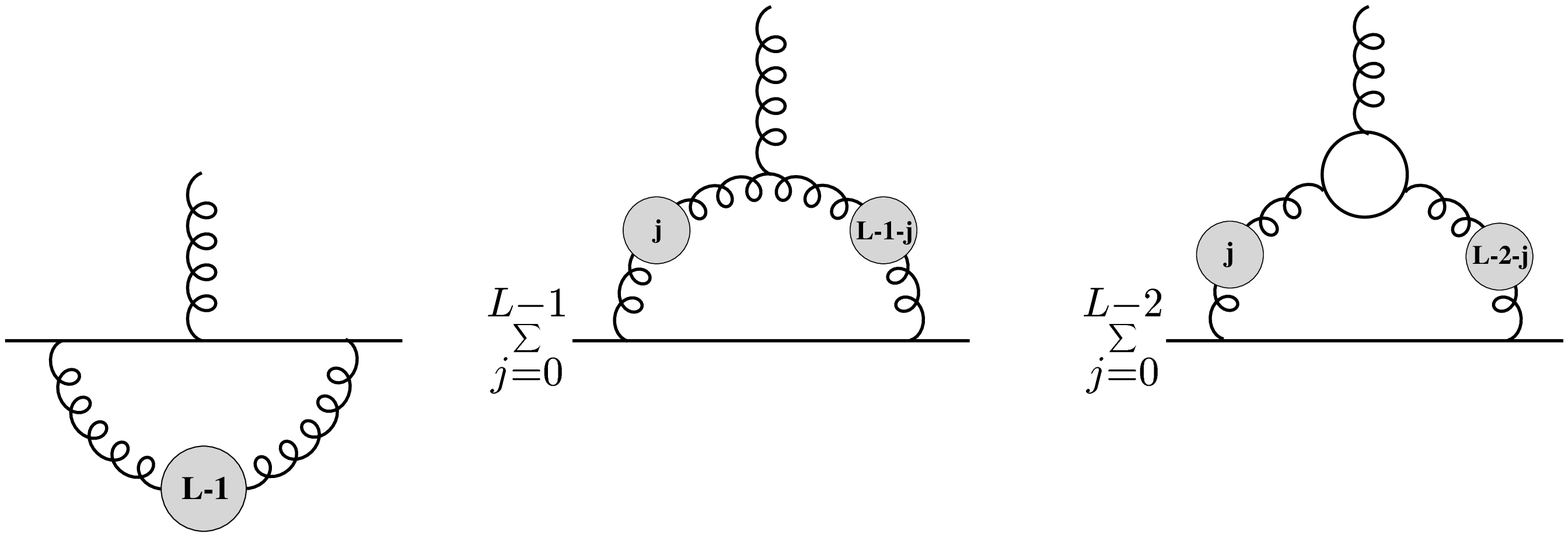}}
\centerline{\parbox{14cm}{\caption{\label{fig:loop}
QCD diagrams for heavy-quark scattering in an external field. A gray
circle with a number $n$ inside represents a chain on $n$ light-quark
loop insertions on a gluon propagator.}}}
\end{figure}

The coefficient $C(m,\mu)$ is obtained by matching scattering
amplitudes of an on-shell heavy quark in an external field in QCD and
HQET, including terms of order $1/m$~\cite{EiHi,CG}. As mentioned
above, our focus is on $L$-loop diagrams with $L-1$ light-quark loops.
All HQET diagrams vanish because they contain no mass scale. The
relevant QCD diagrams are shown in Fig.~\ref{fig:loop}. They must be
supplemented by the wave-function renormalization of the external quark
fields~\cite{BG}. We find that in the \MS scheme the result for the
function $F(\ep,u)$ in~(\ref{Struct}) has the form
\begin{equation}
   F(\ep,u) = \left( \frac{\mu}{m}\right)^{2u} e^{\gamma\ep}\,
   \frac{\Gamma(1+u)\Gamma(1-2u)}{\Gamma(3-u-\ep)}\,
   D(\ep)^{u/\ep-1}\,\Big[ C_F N_F(\ep,u) + C_A N_A(\ep,u) \Big] \,,
\label{Feu}
\end{equation}
where
\begin{equation}
   D(\ep) = 6 e^{\gamma\ep}\,\frac{\Gamma(1+\ep)\Gamma^2(2-\ep)}
   {\Gamma(4-2\ep)} = 1 + \frac{5\ep}{3} + \dots
\label{De}
\end{equation}
is related to the contribution of a light-quark loop to the gluon
self-energy, and
\begin{eqnarray}
   N_F(\ep,u) &=& 4u(1+u-2\ep u) \,, \nonumber\\
   N_A(\ep,u) &=& \frac{2-u-\ep}{2(1-\ep)}\,\left[
    (2-5\ep+2\ep^2) + (3-6\ep+4\ep^2)u \right] \,.
\label{Nres}
\end{eqnarray}
These formulae reproduce the known $L=1$ \cite{EiHi} and $L=2$
\cite{CG} results, where $L=u/\ep$.

Having obtained an explicit result for the function $F(\ep,u)$, we are
ready to derive all-order results for the anomalous dimension and
Wilson coefficient of the chromomagnetic operator in the
large-$\beta_0$ limit. From~(\ref{gamma}), we obtain
\begin{eqnarray}
   \gamma &=& \frac{2 C_A}{\beta_0}\,
    \frac{b(1+2b)\Gamma(5+2b)}{24(1+b)\Gamma^3(2+b)\Gamma(1-b)}
    + O(1/\beta_0^2) \nonumber\\
   &=& \frac{C_A\alpha_s}{2\pi} \left[ 1
    + \frac{13}{6}\,\frac{\beta_0\alpha_s}{4\pi}
    - \frac 12 \left( \frac{\beta_0\alpha_s}{4\pi} \right)^2
    - 2\zeta(3) \left( \frac{\beta_0\alpha_s}{4\pi} \right)^3
    + \dots \right] \,,
\label{gamma1}
\end{eqnarray}
which reproduces correctly the leading (in $\beta_0$) term of the exact
two-loop result~(\ref{gamma2}). The radius of convergence of the
perturbation series in~(\ref{gamma1}) is $\beta_0|\alpha_s|<4\pi$.

Next, setting $\mu=m$ in~(\ref{finite}), we find that the perturbative
expansion of the Wilson coefficient at the scale $\mu=m$ reads
\begin{eqnarray}
   &&C(m,m) = 1 + \frac{\alpha_s(m)}{4\pi}\,\Bigg\{ 2(C_F+C_A)
    + \left[ \frac{25}{3}\,C_F + \left( \frac{299}{36}
    + \frac{\pi^2}{3} \right) C_A \right]
    \frac{\beta_0\alpha_s(m)}{4\pi} \nonumber\\
   &&\quad + \left[ \left( \frac{317}{9} + \frac 43\pi^2 \right)
    C_F + \left( \frac{3535}{162} + \frac{25}{9}\pi^2
    + \frac{14}{3}\zeta(3) \right) C_A \right]
    \left( \frac{\beta_0\alpha_s(m)}{4\pi} \right)^2
    + \dots \Bigg\} \,,
\label{Cm2}
\end{eqnarray}
which again reproduces the leading term of the exact two-loop result~%
\cite{CG}. Unlike the case of the anomalous dimension, here retaining
only the leading term in $1/\beta_0$ gives a reasonable approximation;
for $N=3$ colours, the exact two-loop result is
\begin{equation}
   C(m,m) = 1 + \frac{13}{6}\,\frac{\alpha_s(m)}{\pi}
   + (2.869\beta_0 - 9.761) \left(\frac{\alpha_s(m)}{\pi}\right)^2 \,,
\label{Cm3}
\end{equation}
showing that the sub-leading term is about one third of the leading
one.

Finally, we quote the result for the next-to-leading logarithmic
correction $c(m)$ in~(\ref{hatC}). Expanding the integral in~%
(\ref{delc}) in powers of the coupling constant, we find the expansion
coefficients
\begin{eqnarray}
   c_1 &=& 2 C_F + \frac{25}{6} C_A + O(1/\beta_0) \,, \nonumber\\
   c_2 &=& \left[ \frac{25}{3} C_F + \left( \frac{145}{18}
    + \frac{\pi^2}{3} \right) C_A \right] \beta_0 + O(\beta_0^0) \,,
    \nonumber\\
   c_3 &=& \left[ \left( \frac{317}{9} + \frac 43 \pi^2 \right) C_F
    + \left( \frac{3535}{162} + \frac{25}{9}\pi^2 + 4\zeta(3)
    \right) C_A \right] \beta_0^2 + O(\beta_0) \,.
\end{eqnarray}
The result for $c_1$ reproduces the leading term in~(\ref{c1}). For
$N=3$ colours, the numerical values of the coefficients are $c_1\approx
15.167$, $c_2\approx 45.147\beta_0$, $c_3\approx 226.64\beta_0^2$,
$c_4\approx 1526.0\beta_0^3$, $c_5\approx 13035\beta_0^4$, etc. They
exhibit a (same-sign) factorial divergence $c_{n+1}\sim
(2\beta_0)^n\,n!$, which renders the perturbation series for $c(m)$
divergent and not Borel summable. On the contrary, the coefficients in
the expansion of the anomalous dimension in (\ref{gamma1}) appear to be
well behaved; in fact, this series is rapidly convergent. Although we
do not expect that the higher-order coefficients obtained in the
large-$\beta_0$ limit accurately represent the exact values, we do
trust in the qualitative features of the results. This allows us to
derive an approximation for the next-to-next-to-leading order (NNLO)
correction $c_2$ in~(\ref{hatC}). Expanding~(\ref{RGEsol}) to second
order in $\alpha_s(m)$, and inserting the known expansion coefficients
for the three-loop $\beta$-function~\cite{Tara}, the two-loop anomalous
dimension~\cite{ABN,CG}, and the two-loop matching coefficient
$C(m,m)$~\cite{CG}, we obtain an exact expression for $c_2$ depending
only on the unknown three-loop coefficient of the anomalous dimension,
for which we write $\gamma_2=C_A(-\beta_0^2 + \eta_1\beta_0 + \eta_0)$,
where $\eta_1$ and $\eta_0$ are unknown. This parametrization takes
into account that the anomalous dimension vanishes in the abelian limit
$C_A=0$ \cite{ABN}. We then expand $c_2=d_{-1}\beta_0 +
\sum_{n=0}^4\,d_n(C_A/\beta_0)^n$ and find
\begin{eqnarray}
   d_{-1} &=& \frac{25}{3} C_F + \left( \frac{145}{18}
    + \frac{\pi^2}{3} \right) C_A = \frac{635}{18} + \pi^2 \,,
    \nonumber\\
   d_0 &=& \left( -31 + \frac{20}{3}\pi^2 \right) C_F^2
    + \left( - \frac{23}{24} + \frac 43\pi^2 \right) C_F C_A
    + \left( - \frac{49}{48} - \frac{28}{9}\pi^2 \right) C_A^2
    \nonumber\\
   &&\mbox{}+ \left( \frac{476}{9} - \frac{16}{3}\pi^2 \right) C_F T_F
    + \left( - \frac{298}{27} + \pi^2 \right) C_A T_F
    + \frac{\eta_1}{4}\,C_A \nonumber\\
   &&\mbox{}+ \left( \frac 43\pi^2\ln 2 - 2\zeta(3) \right)
    (C_A^2 + C_F C_A - 6 C_F^2)  \nonumber\\
   &=& - \frac{21353}{432} - \frac{695}{54}\pi^2
    + \frac{28}{9}\pi^2\ln 2 - \frac{14}{3}\zeta(3) + \frac 34\eta_1
    \,, \nonumber\\
   d_1 &=& -\frac 34 C_F^2 + \frac{31}{8} C_F C_A
    - \frac{275}{18} C_A^2 + \frac{\eta_0}{4}
    = -\frac{370}{3} + \frac 14\eta_0 \,, \nonumber\\
   d_2 &=& -\frac{37}{4} C_F^2 + \frac 54 C_F C_A
    + \frac{6491}{144} C_A^2 = \frac{56771}{144} \,, \nonumber\\
   d_3 &=& \frac{55}{2} C_F^2 - \frac{269}{6} C_F C_A
    - \frac{119}{3} C_A^2 = -\frac{4387}{9} \,, \nonumber\\
   d_4 &=& \frac{121}{2} C_F^2 + 77 C_F C_A
    + \frac{49}{2} C_A^2 = \frac{11449}{18} \,,
\end{eqnarray}
where the numerical values refer to $N=3$ colours. Because of the good
convergence of the perturbation series for the anomalous dimension
indicated by our analysis of the large-$\beta_0$ limit, it is
conservative to assume that $|\eta_1\beta_0 + \eta_0|<\beta_0^2$, which
is equivalent to the statement that the true value of the three-loop
anomalous dimension differs from its value in the large-$\beta_0$ limit
by less than 100\%. Under this assumption, we obtain
\begin{equation}
   c_2(n_f=4) = 210.08\pm 6.25 \,, \qquad
   c_2(n_f=3) = 238.04\pm 6.75 \,.
\end{equation}
The uncertainty due to the unknown terms in the three-loop anomalous
dimension is negligible compared with the overall size of the
coefficient. Using this result together with the exact one-loop
coefficient $c_1$ in~(\ref{c1}), we find for $n_f=4$
\begin{equation}
   \widehat C(m) = \big[ \alpha_s(m) \big]^{9/25}\,
   \left[ 1 + 0.672\alpha_s(m) + (1.33\pm 0.04)\alpha_s^2(m)
   + \dots \right] \,.
\end{equation}
In the context of the heavy-quark expansion, this is the most precisely
known Wilson coefficient to date.

\section{Borel summation and renormalon ambiguities}

As a consequence of its divergent behaviour, the perturbation series
for $c(m)$ must be truncated close to its minimal term, and the
perturbative result for the Wilson coefficient is intrinsically
ambiguous. Although we have explored this feature only in the
large-$\beta_0$ limit, it is believed to be of a rather general nature
(see, e.g., \cite{Muel}). In regularization schemes without an explicit
infrared cutoff (such as dimensional regularization with minimal
subtraction), the perturbative calculation of the coefficients $c_n$
involves an integration over all gluon momenta, including long-distance
contributions from soft gluons. High orders in the expansion probe the
region of increasingly smaller gluon momenta, a regime where
perturbation theory is bound to fail. Remarkably, the perturbation
series knows about its deficiency and signals it through the divergent
behaviour of the expansion coefficients. We will now investigate this
phenomenon in more detail.

A convenient tool to study the ambiguities encountered in the attempt
to resum an asymptotic perturbation series is provided by the Borel
image of that series, given by the function $S(u)$ in the integral
representation~(\ref{delc}). Using~(\ref{Feu}), we find that
\begin{equation}
   S(u) = \frac{\Gamma(u)\Gamma(1-2u)}{\Gamma(3-u)}\,\left[
   4u(1+u) C_F + \frac{(2-u)(2+3u)}{2} C_A \right]
   - e^{-5u/3}\,\frac{C_A}{u} \,.
\label{F0u}
\end{equation}
If it existed, the Laplace integral in~(\ref{delc}) would define the
Borel sum of the perturbation series for $c(m)$ in the large-$\beta_0$
limit. However, the presence of singularities along the integration
contour (i.e.\ for positive values of $u$) makes the integral
ill-defined. In the large-$\beta_0$ limit, the function $S(u)$ has pole
singularities at half-integer values of $u$ called infrared
renormalons~\cite{reno1,tHof}. Any attempt to resum perturbation theory
involves an arbitrary choice of how to deal with these singularities. A
measure of the ambiguity in the value of the Borel sum, which is of the
same size as the minimal term in the series, is provided by the residue
of the nearest singularity, in the present case located at $u=1/2$.
Using~(\ref{aV}), we obtain
\begin{equation}
   \Delta[c(m)] = \frac{1}{\beta_0} \left( 2 C_F
   + \frac 74 C_A \right) \frac{\Lambda_{\mathrm{V}}}{m} \,,
\label{delren}
\end{equation}
with $\Lambda_{\mathrm{V}}=e^{5/6}\LMS$. The situation is similar to
the well-known case of the infrared renormalon ambiguity in the
definition of the pole mass of a heavy quark~\cite{BB,BSUV}. The
perturbation series relating the pole mass to a short-distance mass is
also affected by an infrared renormalon at $u=1/2$, and the
corresponding ambiguity is $\Delta m=2C_F\Lambda_{\mathrm{V}}/\beta_0$.
Taking into account that the chromomagnetic operator appears in the
effective Lagrangian~(\ref{Leff}) multiplied by a power of the inverse
pole mass, we find that
\begin{equation}
   \Delta \left[ \frac{\widehat C(m)}{m} \right]
   = \frac{7C_A}{8C_F}\,\frac{\Delta m}{m^2} + O(1/\beta_0^2) \,.
\end{equation}
Hence, the infrared renormalon ambiguity in the product $\widehat
C(m)/m$ is purely non-abelian, and commensurate with the contributions
of higher-dimensional operators in the $1/m$ expansion. Indeed, in all
predictions for physical quantities, the infrared renormalon
ambiguities must cancel against corresponding ambiguities in the
long-distance matrix elements of some higher-dimensional operators.

When the heavy-quark expansion is applied to calculate a physical
quantity, e.g.\ the mass splitting in (\ref{split}), the resulting
expressions contain short-distance Wilson coefficients and
long-distance hadronic matrix elements. As mentioned above, in
regularization schemes without a hard momentum cutoff the Wilson
coefficients also contain contributions from large distances, where
perturbation theory is ill-defined, and these contributions produce
infrared renormalon ambiguities. Likewise, the hadronic matrix elements
(which are not calculable perturbatively) contain contributions from
small distances, which lead to ultraviolet renormalon ambiguities. In
other words, in such schemes the separation of short- and long-distance
contributions into Wilson coefficients and matrix elements is
intrinsically ambiguous. Only when all contributions are combined to
form a physical quantity, an unambiguous result is obtained. In the
context of the HQET, the cancelations between infrared and ultraviolet
renormalon ambiguities have been traced in detail in~\cite{NS}. Here we
consider the particular case of the mass splitting.

\begin{figure}
\epsfxsize=12cm
\centerline{\epsffile{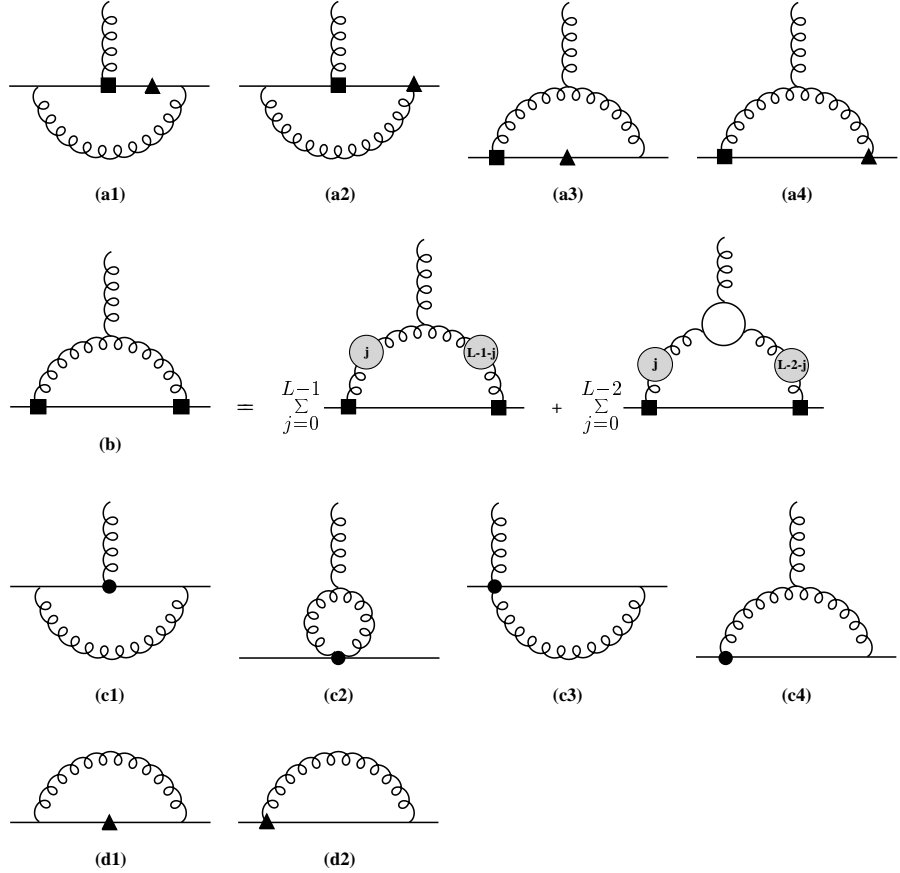}}
\centerline{\parbox{14cm}{\caption{\label{fig:UV}
Diagrams contributing to the ultraviolet renormalons contained in the
parameters $\rho_{\pi G}^3$ (a), $\rho_A^3$ (b), and $\rho_{LS}^3$ (c).
Insertions of the chromomagnetic, kinetic and spin-orbit operators are
indicated by squares, triangles and circles, respectively. It is
implied that light-quark loops are inserted in all possible ways, for
example as shown in the case of diagram (b). Also implied are mirror
copies of the diagrams if appropriate.}}}
\end{figure}

The ultraviolet contributions to the hadronic parameters $\rho_i^3$ are
independent of the nature of the external states and thus may be
calculated using quark and gluon rather than hadron states. At one-loop
order, the relevant diagrams are shown in Fig.~\ref{fig:UV}. On
dimensional grounds, they are linearly divergent in the ultraviolet
region. In dimensional regularization such power divergences are not
seen (by definition), but they reflect themselves in a factorial
divergence of the perturbation series when higher-order diagrams are
taken into account. We shall again analyse these contributions in the
large-$\beta_0$ limit, by inserting an arbitrary number of light-quark
loops in all possible ways into the diagrams of Fig.~\ref{fig:UV}. The
relevant matrix elements are then expanded to linear order in the gluon
momentum $q$ and projected onto the structure of the chromomagnetic
operator. The external quarks are kept off-shell in order to provide
for an infrared regulator. The Borel transforms of these matrix
elements contain ultraviolet renormalon poles at $u=1/2$. For
dimensional reasons, the residues are proportional to
$\Lambda_{\mathrm{V}}$ times the matrix element $\mu_G^2$ of the
chromomagnetic operator, which is the only lower-dimensional operator
contributing to the mass splitting. Therefore, the leading ultraviolet
renormalon ambiguities of the $\rho_i^3$ parameters are proportional to
$\mu_G^2\Delta m$. Note that diagrams in which the external gluon is
attached to a heavy-quark line, or to a two-gluon kinetic or
chromomagnetic vertex, do not contribute. The diagram (c1) does not
contribute to the chromomagnetic structure since the spin-orbit vertex
only couples to the chromoelectric field. Moreover, with a suitable
choice of the external momenta one can achieve that the diagrams (c2)
and (c3) vanish. (Their sum vanishes with any choice of momenta.) A
subtlety that needs to be taken into account is that there exists an
ultraviolet renormalon at $u=1/2$ in the $1/m$ suppressed contribution
to the quark wave-function renormalization with a kinetic-energy
insertion, shown in diagrams (d1) and (d2); the corresponding ambiguity
$\Delta Z_Q=\frac 32\Delta m/m$ in the normalization of $\mu_G^2$
contributes to $\Delta\rho_{\pi G}^3$ and cancels the $C_F$ terms in
the diagrams (a1) and (a2). From a direct evaluation of the
non-vanishing diagrams, using the method described above, we obtain for
the ultraviolet renormalon ambiguities in the large-$\beta_0$ limit
$\Delta\rho_i^3 = -k_i (C_A/C_F) \mu_G^2\Delta m$ with coefficients
$k_{\pi G}=2/3$, $k_A=19/12$, and $k_{LS}=1/2$. The sum of the
ultraviolet ambiguities in~(\ref{split}) is thus given by
\begin{equation}
   \Delta\rho_{\pi G}^3 + \Delta\rho_A^3 - \Delta\rho_{LS}^3
   = - \frac{7C_A}{4C_F}\,\mu_G^2\,\Delta m + O(1/\beta_0^2) \,.
\end{equation}
It precisely cancels the infrared renormalon ambiguity in the leading
term.

It is interesting that the requirement of a cancelation of renormalon
ambiguities in physical quantities allows us to derive further
information about the asymptotic behaviour of the expansion
coefficients $c_n$ in~(\ref{hatC}) without any additional dynamical
input~\cite{reno2,Bene}. The point is that, beyond the large-$\beta_0$
limit, the different terms in~(\ref{MVMP}) contain different powers of
$[\alpha_s(m)]^{\gamma_0/2\beta_0}$, since they contain different
powers of the Wilson coefficient $\widehat C(m)$. These leading
logarithms are exactly known to all orders of perturbation theory. They
multiply the renormalon ambiguities of the various terms. In order to
maintain the cancelation between infrared and ultraviolet renormalon
ambiguities in the presence of the leading logarithms, the Borel
transform for the Wilson coefficient in~(\ref{F0u}) has to be modified.
In the vicinity of $u=1/2$, the simple pole in
\begin{equation}
   S(u) = \left( 2 C_F + \frac 74 C_A \right)
   \frac{1}{\frac12 - u} + \dots
\end{equation}
must be replaced by a sum of branch points,
\begin{equation}
   S(u) = \frac{1}{\big(\frac12 - u\big)^{1+\beta_1/2\beta_0^2}}
   \left[ 2 C_F K_1 - \frac13 C_A K_2 + \frac{19}{12}\,
   \frac{C_A K_3}{\big(\frac12 - u\big)^{-\gamma_0/2\beta_0}}
   + \frac12\,\frac{C_A K_4}{\big(\frac12 - u\big)^{\gamma_0/2\beta_0}}
   \right] \,,
\end{equation}
where $\gamma_0=2 C_A$ is the one-loop coefficient of the anomalous
dimension of the chromomagnetic operator given in (\ref{gamma2}), and
$\beta_1$ denotes the two-loop coefficient of the $\beta$-function. The
normalization constants $K_i=1+O(1/\beta_0)$ are undetermined beyond
the large-$\beta_0$ limit \cite{Bene}. Up to corrections of order
$1/n$, the corresponding asymptotic behaviour of the expansion
coefficients $c_n$ is
\begin{equation}
   c_{n+1} = (2\beta_0)^n\,n!\,n^{\beta_1/2\beta_0^2}
   \left[ 4 C_F K_1 - \frac 23 C_A K_2
   + \frac{19}{6} C_A K_3\,n^{-\gamma_0/2\beta_0}
   + C_A K_4\,n^{\gamma_0/2\beta_0} \right] \,.
\end{equation}
For very large $n$, the last term gives the dominant behaviour (since
$\gamma_0/\beta_0>0$); however, for moderate values of $n$ all
contributions are of similar importance.

\begin{figure}
\epsfxsize=10cm
\centerline{\epsffile{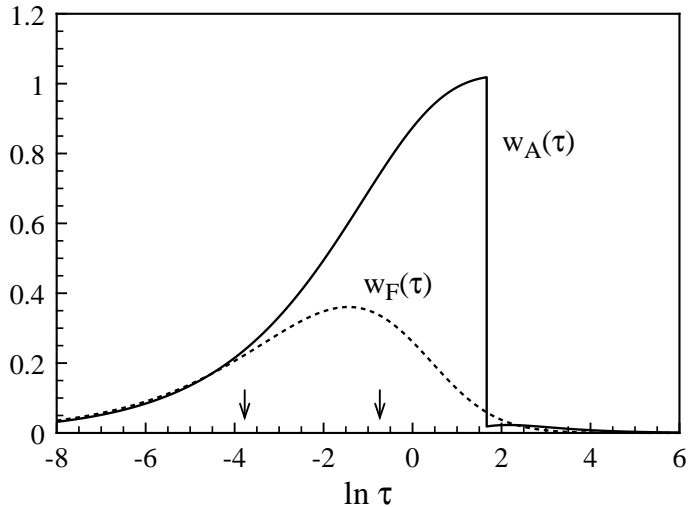}}
\centerline{\parbox{14cm}{\caption{\label{fig:plot}
Distribution functions $w_F(\tau)$ (dashed) and $w_A(\tau)$ (solid).
The arrows show the factorization point for $\lambda=1\,$GeV and
$m=4.8\,$GeV (left), $m=1.44\,$GeV (right).}}}
\end{figure}

In order to evaluate the Borel integral (\ref{delc}), it is useful to
rewrite it in the form~\cite{scale}
\begin{eqnarray}
   c(m) &=& \frac{1}{\beta_0} \int\limits_{-\infty}^\infty\!
    d\ln\tau\,\frac{w(\tau)}{\ln\tau + \ln\big(m/\Lambda_V\big)^2}
    + O(1/\beta_0^2) \nonumber\\
   &=& \int\limits_0^\infty\frac{d\tau}{\tau}\,
    w(\tau)\,\frac{\alpha_s(\sqrt{\tau}e^{-5/6}m)}{4\pi}
    + O(1/\beta_0^2) \,.
\label{distr}
\end{eqnarray}
The function
\begin{equation}
   w(\tau) = \frac{1}{2\pi i} \int\limits_{u_0-i\infty}^{u_0+i\infty}
   \!du\,S(u)\,\tau^u = C_F w_F(\tau) + C_A w_A(\tau) \,;\quad
   0 < u_0 < \textstyle{\frac 12} \,, 
\end{equation}
which is the inverse Mellin transform of the Borel image $S(u)$,
describes the distribution of gluon virtualities in the one-loop
diagrams in Fig.~\ref{fig:loop}, which contribute to the calculation of
$c(m)$. Using the methods developed in~\cite{scale}, we obtain
\begin{eqnarray}
   w_F(\tau) &=& \frac{4}{\sqrt{1+4/\tau}} - 4\tau
    + 2\tau^2 \left( \sqrt{1+4/\tau} - 1 \right) \,, \nonumber\\
   w_A(\tau) &=& \frac{5\tau}{4} \left( \sqrt{1+4/\tau} - 1 \right)
    - \frac{3}{2\sqrt{1+4/\tau}} - \Theta(\tau-e^{5/3}) \,.
\end{eqnarray}
These functions are shown in Fig.~\ref{fig:plot}. Note that the
integral in (\ref{distr}) runs over the Landau pole in the running
coupling constant, located at $\tau_L=(\Lambda_{\mathrm{V}}/m)^2$. This
is how infrared renormalons make their appearance. As in the case of
the original Borel integral, we must specify an arbitrary prescription
of how to deal with the Landau singularity. The renormalon ambiguity is
given by the residue of the pole, i.e.\ $\Delta[c(m)] =
w(\tau_L)/\beta_0$. From an expansion of the distribution function in
the region $\tau\ll 1$, we readily recover the result (\ref{delren}).

The representation~(\ref{distr}) makes explicit that perturbative
calculations contain long-distance contributions from the region of low
momenta in Feynman diagrams. At the same time, it provides for a
convenient way to separate these long-distance contributions from the
short-distance ones by introducing a hard separation scale $\lambda$.
We may then define the short-distance coefficient
$c_{\mathrm{sd}}(m,\lambda)$ as
\begin{equation}
   c_{\mathrm{sd}}(m,\lambda) = \int\limits_{\lambda^2}^\infty
   \frac{\hbox{d}\mu^2}{\mu^2}\, w(\mu^2/m^2)\,
   \frac{\alpha_s(e^{-5/6}\mu)}{4\pi} + O(1/\beta_0^2) \,.
\label{csd}
\end{equation}
As long as $\lambda$ is chosen large enough, this coefficient can be
reliably calculated in perturbation theory and is free of renormalon
ambiguities. The long-distance contribution eliminated by this
procedure must be combined with the non-perturbative power corrections
in the heavy-quark expansion.

\section{Numerical results and conclusions}

Our goal was to study the large-order behaviour of the
Wilson coefficient of the chromomagnetic operator in the HQET
Lagrangian. The renormalization-group invariant coefficient $\widehat
C(m)$ defined in (\ref{hatC}) is known exactly at next-to-leading order
\cite{ABN}. Here, we have derived an excellent next-to-next-to-leading
order approximation by combining the exact two-loop matching condition
obtained in \cite{CG} with an approximation for the three-loop
anomalous dimension. This estimate has been obtained by summing
fermion-loop contributions to all orders in perturbation theory, thus
deriving the leading term in an expansion in powers of $1/\beta_0$. We
have also derived an all-order result for the Wilson coefficient in the
large-$\beta_0$ limit, finding that the perturbation series for
$\widehat C(m)$ exhibits a divergent behaviour already in low orders,
caused by a nearby infrared renormalon singularity. The resulting
ambiguity is commensurate with terms of order $1/m^2$ in the effective
Lagrangian, whose corresponding ultraviolet renormalons we have
identified.

Let us now investigate the implications of our results by comparing the
various approximations for the Wilson coefficient. To this end, we
evaluate the partial sums
\begin{equation}
   \widehat C_{(N)}(m) = \big[ \alpha_s(m) \big]^{\gamma_0/2\beta_0}\,
   \left[ 1 + \sum_{n=1}^N c_n \left( \frac{\alpha_s(m)}{4\pi}
   \right)^n \right]
\end{equation}
for $n_f=4$ light quark flavours, using $\alpha_s(m_b)=0.22$ and
$\alpha_s(m_c)=0.36$ in the \MS scheme. The ratio of the two
coefficient functions at these scales (for $n_f=4$) gives the
perturbative correction to the ratio of mass splitting between the
ground-state pseudoscalar and vector mesons in the bottom and charm
systems~\cite{ABN}:
\begin{equation}
   \frac{M_{B^*}^2-M_B^2}{M_{D^*}^2-M_D^2}
   = \frac{\widehat C(m_b)}{\widehat C(m_c)}\,\left[ 1
   + \Lambda_{\mathrm{eff}} \left( \frac{1}{m_c} - \frac{1}{m_b}
   \right) + \dots \right] \approx 0.89 \,,
\label{splitratio}
\end{equation}
where $\Lambda_{\mathrm{eff}}$ is a combination of the hadronic
parameters $\rho_i^3$ introduced in (\ref{split}), and for simplicity
we neglect short-distance corrections in the $1/m$ terms.

\begin{table}
\caption{\label{tab:1}
Perturbative approximations for the Wilson coefficients $\widehat
C_{(N)}(m_b)$ and $\widehat C_{(N)}(m_c)$}
\vspace{0.4cm}
\begin{center}
\begin{tabular}{|c|cc|cc|cc|}\hline
\rule[-0.35cm]{0cm}{0.95cm}
 & \multicolumn{2}{c|}{$\widehat C_{(N)}(m_b)$} &
 \multicolumn{2}{c|}{$\widehat C_{(N)}(m_c)$} &
 \multicolumn{2}{c|}{$\widehat C_{(N)}(m_b)/\widehat C_{(N)}(m_c)$} \\
 $N$ & large-$\beta_0$ & exact & large-$\beta_0$ & exact &
 large-$\beta_0$ & exact\\[0.15cm]
\hline
\rule{0cm}{0.5cm} 0 \rule{0cm}{0.5cm}
   & 0.580 & 0.580    & 0.692 & 0.692    & 0.838 & 0.838    \\[0.15cm]
 1 & 0.734 & 0.659    & 0.993 & 0.847    & 0.739 & 0.778    \\[0.15cm]
 2 & 0.801 & 0.697(1) & 1.207 & 0.967(4) & 0.664 & 0.721(3) \\[0.15cm]
 3 & 0.850 &          & 1.463 &          & 0.581 &          \\[0.15cm]
 4 & 0.898 &          & 1.875 &          & 0.479 &          \\[0.15cm]
 5 & 0.958 &          & 2.714 &          & 0.353 &          \\[0.15cm]
\hline
\rule{0cm}{0.5cm} Borel sum \rule{0cm}{0.5cm}
              & 0.839(38) & & 1.027(136) & &
 $0.817_{-0.065}^{+0.082}$ & \\[0.15cm]
 min.\ term   & 0.898(48) & & 1.207(214) & &
 $0.744_{-0.078}^{+0.112}$ & \\[0.15cm]
 $\mu>1\,$GeV & 0.774     & & 0.906      & & 0.854      & \\[0.15cm]
\hline
\end{tabular}
\end{center}
\end{table}

Our predictions for the Wilson coefficients $\widehat C_{(N)}(m_b)$ and
$\widehat C_{(N)}(m_c)$ are given in Table~\ref{tab:1}. In the upper
part of the Table, we show the partial sums obtained at different orders
in perturbation theory, both in the large-$\beta_0$ limit and exactly,
so far as the exact results are known. In the large-$\beta_0$ limit,
the divergent behaviour of the perturbation series sets in already in
low orders. At the scale $m_b$, the minimal term in the series is
reached around $N=4$, whereas it is reached around $N=2$ at the scale
$m_c$. As a consequence, the values of the coefficient functions at the
two scales become more and more different for larger $N$, and their
ratio drifts away from the experimental value of the ratio of mass
splittings.

The reason for this behaviour lies in the nearby location of the first
infrared renormalon. This can be understood intuitively by considering
the distribution functions shown in Fig.~\ref{fig:plot}, which are
broad and extend far into the infrared region. Note that the point
$\mu=m$ corresponds to $\ln\tau=5/3$ in the \MS scheme. Hence, there
are essentially no contributions from scales above $m$ (in the case of
the $C_A$ terms those are subtracted by \MS renormalization), but on
the other hand the distribution functions decrease very slowly
($\sim\sqrt\tau$) in the infrared region $\tau\to 0$. The strength of
this decrease is determined by the position of the nearest infrared
renormalon singularity. In general, if the nearest infrared renormalon
is located at $u=k$, then $w(\tau)\sim\tau^k$ for $\tau\to 0$
\cite{scale}. As a consequence, the dominant contributions come from
scales significantly below the heavy-quark mass.\footnote{For the
combined distribution function $w(\tau)$ the average value of $\ln\tau$
is $-1.310$, corresponding to the scale $\mu_{\mathrm{BLM}}\approx
0.226 m$ in the \MS scheme. This is precisely the scale obtained in the
BLM scale-setting prescription~\protect\cite{BLM}. In view of the low
value of the average virtuality, the bad convergence of the
perturbation series is not surprising.}
In the lower portion of Table~\ref{tab:1}, we show the results for the
Wilson coefficients obtained by taking the principal value of the Borel
integral. The quoted errors reflect the renormalon ambiguities, which
are sizeable, in particular, at the scale $m_c$. The Borel resummation
gives values close to those obtained when the series is truncated at
the minimal term (in the latter case the error is taken to be the size
of the minimal term), as it is expected on general grounds. In the last
row we show the short-distance contributions to the coefficient
functions arising from virtualities above $1\,$GeV (in the V scheme),
as defined in (\ref{csd}). The portion of infrared contributions is
much larger at the scale $m_c$ than it is at the scale $m_b$. This is
clearly seen in Fig.~\ref{fig:plot}, where the arrows indicate the
location of the separation scale $\lambda=1\,$GeV (corresponding to 435
MeV in the \MS scheme). If the infrared contributions are cut away, the
resulting values of the coefficients are again close to their values
obtained from low-order calculations.

Although the all-order results are instructive, we must not forget that
they are obtained in a very questionable approximation scheme (the
large-$\beta_0$ limit). Indeed, the few exact results available
indicate that the perturbation series may be much better behaved than
indicated by the large-$\beta_0$ limit. Still, we trust in the
qualitative observations that the series start to diverge at some (low)
order, and that the onset of the divergence is reached early the lower
the scale $m$ is. We will now explore the implications of these results
for the phenomenology of the heavy-quark expansion. For a consistent
inclusion of power-suppressed effects the perturbative coefficients of
the leading terms must be known with sufficient accuracy. Different
truncation or resummation schemes imposed on the short-distance
coefficients imply different definitions of the hadronic parameters
appearing at higher order in the heavy-quark expansion, such as the
parameter $\Lambda_{\mathrm{eff}}$ in (\ref{splitratio}). Only when the
perturbative coefficients are truncated close to their asymptotic
value, the hadronic parameters are $m$ independent as they should. When
applied to the particular case of the chromomagnetic operator, the
conclusion is that there is not much to be gained by calculating
$\widehat C(m)$ beyond the first few orders of perturbation theory.
Taking into account that the series for $\widehat C(m_c)$ diverges
earlier than that for $\widehat C(m_b)$, we may argue that the optimal
perturbative prediction for the ratio based on exact information is
obtained by combining the NNLO result at the bottom scale with the NLO
result at the charm scale. This gives
\begin{equation}
   \frac{\widehat C_{(2)}(m_b)}{\widehat C_{(1)}(m_c)} \approx 0.80 \,.
\label{opt}
\end{equation}
In the large-$\beta_0$ limit, the corresponding ratio equals 0.81 and
is indeed very close to the principal value of the Borel integral.
Using the result (\ref{opt}), we conclude that the power corrections in
(\ref{splitratio}) give a contribution of about 11\%, corresponding to
a scale $\Lambda_{\mathrm{eff}}\approx 220\,$MeV. Thus, if the
asymptotic behaviour of perturbation theory is carefully taken into
account, it appears that power corrections in the heavy-quark expansion
are well under control.

\vspace{0.3cm}
{\it Acknowledgments:\/}
We are grateful to M.~Beneke for useful discussions. A.G.G.\ thanks the
CERN Theory Division for its hospitality during the main part of this
work.

\end{document}